\documentclass[3p]{elsarticle}
\usepackage[english]{babel}
\usepackage[utf8]{inputenc}
\usepackage[OT4]{fontenc}
\usepackage{hyperref}
\usepackage{amsmath}
\usepackage{amsthm}
\usepackage{graphicx}
\usepackage{caption}
\usepackage{subcaption}
\usepackage{amsthm}
\usepackage{amssymb}
\usepackage{multirow}
\usepackage{rotating}
\usepackage{booktabs}

\newcommand{\R}{\mathbb{R}}

\newcommand{\phase}{\mathbb{H}}

\newcommand{\W}{\mathcal{W}}

\newcommand{\param}{\mathbb{P}}
\newcommand{\rr}[2]{\frac{\partial #1}{\partial #2}}
\newcommand{\pr}[1]{\frac{\partial}{\partial #1}}

\newcommand{\post}{\text{post}}
\newcommand{\pre}{\text{pre}}
\newcommand{\eq}{\text{eq}}

\providecommand{\sci}[2]{{\begingroup{}\setlength{\medmuskip}{0mu}$#1\times{}10^#2$\endgroup{}}}

\newtheorem{lemma}{Lemma}
\newtheorem{thm}{Theorem}

\begin{document}
\begin{frontmatter}
\title{Adjoint Lattice Boltzmann for Topology Optimization on multi-GPU architecture}
 
 \author{\L{}.~{\L}aniewski-Wo\l{}\l{}k}
%\fnref{fn1}
 \ead{llaniewski@meil.pw.edu.pl}
 
 \author{J.~Rokicki}
 \ead{jack@meil.pw.edu.pl}
  
 %\fntext[fn1]{This is the specimen author footnote.}
 
\address{Institute of Aeronautics and Applied Mechanics, Warsaw University of Technology, Nowowiejska 24, 00-665 Warszawa, Poland}
 
\begin{abstract}
In this paper we present a topology optimization technique applicable to a broad range of flow design problems. We propose also a discrete adjoint formulation effective for a wide class of Lattice Boltzmann Methods (LBM). This adjoint formulation is used to calculate sensitivity of the LBM solution to several type of parameters, both global and local. The numerical scheme for solving the adjoint problem has many properties of the original system, including locality and explicit time-stepping. Thus it is possible to integrate it with the standard LBM solver, allowing for straightforward and efficient parallelization (overcoming limitations typical for discrete adjoint solvers). This approach is successfully used for the channel flow to design a free-topology mixer and a heat exchanger. Both resulting geometries being very complex maximize their objective functions, while keeping viscous losses at acceptable level.
\end{abstract}
\end{frontmatter}

\section{Introduction}
The goal of the present work was to develop an optimization framework for a variety of multi-physics fluid flow problems. This framework is intended to provide consistent objective function improvement for any physical model implemented, while the numerical method is expected to allow for straightforward parallelization to achieve peak performance on clusters of modern nVidia CUDA capable GPUs.

Optimal design in fluid flow problems has received considerable attention in recent years by both engineers and mathematicians. For a survey of CFD optimization methods we refer the reader to the book by Mohammadi and Pironneau~\cite{mohammadi_applied_2001}. In general, flow domain optimization problems are solved with two main approaches called shape and topology optimization.

The first, more traditional approach, consists in parametrization of the geometry and subsequent finding the parameters corresponding to the best objective function~\cite{jameson_aerodynamic_2003}. This limits the range of available shapes as topology cannot change during the optimization process. Nevertheless the approach remains very efficient and accurate in assessing the influence of fine details of geometry.

The second approach defines the unknown shape using a global scalar field indicating presence of fluid or solid at each point of the domain (different scalar fields are used for this purpose, nevertheless the artificial porosity seems to be most popular~\cite{othmer_continuous_2008,andreasen_topology_2009}). This approach is very well established for solids and structures; for an extensive review we refer the reader to the book by Bends\o{}e and Sigmund~\cite{bendsoe_topology_2003}.

Discretization of a fluid problem is typically obtained by finite volume, finite element, or discontinuous Galerkin methods. These methods use fine body-fitted meshes, which may be difficult to generate for complex 3D geometries. Moreover these approaches are quite difficult to apply for multiphase flows with variable complex interfaces. In contrast Lattice Boltzmann Methods (LBM) are based on the concept of solving discrete Boltzmann equation on Cartesian grids. This in principle allows to tackle geometries of extreme complexity (e.g., see application of LBM to the filtration problems~\cite{pan_evaluation_2006,regulski_flows_2013}). LBM is proven to converge to incompressible Navier-Stokes equations for the low Mach number limit and in recent years has gained broad recognition for calculation of micro-fluids, multi-phase problems, and flows through porous media~\cite{lee_stable_2005,pan_evaluation_2006,regulski_flows_2013}. For a comprehensive introduction to LBM we refer the reader to the book by Succi~\cite{succi_lattice_2001}. The geometry in LBM is described by switching specific nodes of the mesh on and off and applying a so-called Bounce Back boundary condition. This inherent immersed boundary approach is an ideal candidate for topology optimization technique, where very complex shapes are expected to appear.

To achieve acceptable efficiency of the optimization algorithm for high-dimensional design spaces, the gradient of the objective function is needed. For this purpose we use a well established adjoint approach~\cite{othmer_continuous_2008,pingen_topology_2007,andreasen_topology_2009,mohammadi_applied_2001}. In the literature, two main types of this approach can be found. The first is a continuous adjoint, which bases on equation dual to the original form of the problem (e.g., Navier-Stokes equations). The resulting dual equation is discretized, in most cases, by the same scheme as the original (primal) problem~\cite{othmer_continuous_2008}. The second approach is a discrete adjoint, which formulates equations dual to the already discretized form of the original problem~\cite{jaworski_toward_2008,muller_performance_2005}. In the present work we use the latter approach, as it can be easily automated, at the same time providing the accurate value of the computed gradient.

Despite advantages, the discrete adjoint exhibits drawbacks related to parallelization. This is because many implementations use Automatic Differentiation (AD) techniques to the whole residual calculation procedure. In most cases AD tools are incapable of handling MPI directives\footnote{More modern AD software can in principle cope with MPI directives~\cite{utke_toward_2009}, but the process is still not completely automated and may produce a very inefficient code.}, which makes the resulting code inherently serial. In contrast, in the present work we apply AD only to small local procedures, which allows the adjoint code to be both efficient and parallel.

The optimization in the context of Lattice Boltzmann Methods was addressed previously by Pingen, Evgrafov et al.~\cite{pingen_topology_2007,pingen_adjoint_2009,makhija_topology_2012} also with the use of adjoint approach. The adjoint for Lattice Boltzmann Methods was also considered for parameter identification by Tekitek et al.~\cite{tekitek_adjoint_2006}, but not for geometry optimization. The problem of low-Reynolds number mixing was addressed earlier with both Finite Element Method by Andreasen et al.~\cite{andreasen_topology_2009}, and with LBM by Makhija et al.~\cite{makhija_topology_2012}, however the results of the latter work are limited to the 2D channel flows and represent low topological complexity. In all of the above cases, the adjoint formulation is derived by hand for a specific model.

In contrast the present work provides a fully parallel adjoint formulation for a wide class of Lattice Boltzmann schemes, and applies the method to 3D topological optimization in the context of the coupled flow and heat transfer problem. The optimization is taking advantage of a one-shot approach, which further lowers the numerical cost. The obtained 3D optimal heat exchanger has complex, almost fractal geometry.

In Section~\ref{sec:framework} we describe a framework for a class of numerical LBM schemes and provide for them an adjoint formulation. In Section~\ref{sec:lb} we show details of the current LBM presenting: (i) the topology parametrization, (ii) the optimization algorithm and (iii) the implementation for the CUDA architecture. In Section~\ref{sec:results} we formulate two distinct topology optimization problems and analyze the results obtained. We also investigate the parallel performance of the code, considering both weak and strong scaling. The two test cases concern: a low Raynolds number mixer and a free-topology heat exchanger. The final Section presents the conclusions of the presented work.

\section{Local Lattice Boltzmann Scheme and adjoint formulation}\label{sec:framework}
To address a wider class of Lattice Boltzmann Methods and to preserve important parallelization properties in the adjoint formulation, we define a subclass of Local Lattice Boltzmann Schemes (LLBS). The procedure of constructing an adjoint formulation presented in this paper can be applied to any Lattice Boltzmann Method that fits into the proposed subclass.

The main features of LLBS are locality of the collision operator and reversibility of the streaming. These constraints still allow to have a different collision operator for each lattice node, provided that the operator does not use information from the surrounding nodes (apart from the streamed information). Most Lattice Boltzmann schemes, including single and multiple relaxation time schemes, fit into this framework. Many other schemes, including those for multi-phase flows, can be suitably reformulated for this framework by adding additional densities. These properties of LLBS allow for a unified formulation of the adjoint problem, at the same time preserving high parallel efficiency for different physical models.

\subsection{Local Lattice Boltzmann Scheme}

For the purpose of generalization we propose the following mathematical framework. Let us define lattice $L$ as a finite set $L=\{0,\ldots,L_x-1\}\times\{0,\ldots,L_y-1\}\times\{0,\ldots,L_z-1\}$. For a fixed $M$ we define density space $\R^M$, with its elements describing the physical state at each node. The complete state of our system is expressed by the states at all the lattice nodes $f:L\to \R^M$. The linear space of all possible states will be called a state space $\phase = \{f:L\to \R^M\}$. Let us also define a parameter space $\param\subset\R^p$ describing the allowable range for the optimization variables $\alpha\in\param$.

The Lattice Boltzmann Method is characterized by complete separation of communication between nodes (called streaming) and calculations at each node (called collision). This feature allows for very efficient massive parallelization and nearly linear scalability~\cite{pohl_performance_2004}.

At each lattice node $x\in L$, the streaming is defined as:
\[(Sf)_j(x+e_j) = f_j(x)\quad\text{for }j=1,\cdots,M\]
where $e_j$ denotes the shift of information related to the $j$-th density from the node $x$ to the neighbouring node $x+e_j$. The shift $x+e_j$ is assumed to be periodic in all directions (further on, the increment/decrement shift will always mean addition/subtraction modulo $L_x$, $L_y$ and $L_z$). This periodicity of the streaming operator is vital for simplicity of the adjoint formulation, yet by means limits the generality of the approach.

The collision operator $W^x(f(x),\alpha)$ is defined for each node $x$ of the Lattice, its first argument denotes the set of densities, while the second the possible dependence on all optimization parameters $\alpha\in\param$.
In general we may have a separate collision operator $W^x:\R^M\to \R^M$ for each node of the lattice $x\in L$.

A single step $n$ of the Local Lattice Boltzmann iterative scheme is finally defined as a composition of streaming and collision operators:
\begin{equation}f^{n+1}_j(x+e_j) = \W^x_j(f^n(x),\alpha)\quad\text{for }j=1,\cdots,M\label{eq:normal:iteration}\end{equation}

It has to be stressed again, that $\W^x$ depends only on the densities at a given node, but not on the densities at the neighbouring nodes.

The stationary solution $\hat{f}$, forms a fixed point of the iterative scheme~\eqref{eq:normal:iteration}:
\begin{equation}\hat{f}_j(x+e_j) = \W^x_j(\hat{f}(x),\alpha)\label{eq:normal:stationary}\end{equation}
It should be noted that this stationary solution $\hat{f}$ implicitly depends on all parameters $\alpha$.

It will be assumed now, that the objective function, we want to optimize, has the following general form:
\begin{equation}F(\alpha)=\sum_x F^x(\hat{f}(x))\label{eq:objective:function}\end{equation}
where node dependent function $F^x$ can be quite arbitrary. Other forms of~\eqref{eq:objective:function} are possible, yet this formula seems sufficient to represent typical optimization problems in hydrodynamics.

\subsection{Adjoint Lattice Boltzmann Scheme}

Before we go any further, we need to make a simple observation, based on periodicity of the shift operator $S$.
\begin{lemma}\label{lem:tautologia}
For any $f\in\phase$, we have $\sum_{x,i} f_i(x) = \sum_{x,i} f_i(x+e_i)$. From which it follows that for any $a,b\in\phase$:
\[\sum_{x,i} a_i(x)b_i(x) = \sum_{x,i} a_i(x+e_i)b_i(x+e_i)\quad\text{and}\quad\sum_{x,i} a_i(x)b_i(x+e_i) = \sum_{x,i} a_i(x-e_i)b_i(x)\]
\end{lemma}
It can be noted that, for a scalar product $\langle a,b\rangle = \sum_{x,i}a_i(x)b_i(x)$ defined on $H$, this lemma states that the streaming operator $S$ is unitary, and we have both $\langle a,b\rangle = \langle Sa,Sb\rangle$ and $\langle a,Sb\rangle = \langle S^{-1}a,b\rangle$.

Now we can introduce the adjoint Lattice Boltzmann equation and prove the required properties.
\begin{thm}
If $\hat{f}\in\phase$ is a fixed point solution of the Local Lattice Boltzmann equation~\eqref{eq:normal:stationary}, the objective function $F$ is defined by~\eqref{eq:objective:function} and $v\in\phase$ is a solution of the equation:
\begin{equation}v_k(x-e_k) = \sum_{j}v_j(x)\rr{\W^x_j}{f_k}(\hat{f}(x),\alpha) + \rr{F^x}{f_k}(\hat{f}(x))\quad\text{for }k=1,\cdots,M\label{eq:adjoint:stationary}\end{equation}
then the gradient of the objective function can be calculated as
\begin{equation}\rr{F}{\alpha} = \sum_{x,j} v_j(x)\rr{\W^x_j}{\alpha}(\hat{f}(x),\alpha)\label{eq:adjoint:derivative}\end{equation}
\end{thm}

By~\eqref{eq:adjoint:stationary} we have defined the fixed-point adjoint Lattice Boltzmann equation. It is important to notice, that $v$ is of the same size as $\hat{f}$, while the number of equations in~\eqref{eq:adjoint:stationary} is independent of the number of parameters $\alpha$. Solved once, it can be used to calculate gradient of the objective function with respect to any number (and any type) of parameters by using the formula~\eqref{eq:adjoint:derivative}.

\begin{proof}
Let us differentiate equation~\eqref{eq:normal:stationary} with respect to $\alpha$:
\begin{equation}\rr{\hat{f}_j}{\alpha}(x+e_j) = \sum_{k}\rr{\W^x_j}{f_k}(\hat{f}(x),\alpha)\rr{\hat{f}_k}{\alpha}(x) + \rr{\W^x_j}{\alpha}(\hat{f}(x),\alpha)\label{eq:normal:tangent}\end{equation}
Now let us multipy equation~\eqref{eq:adjoint:stationary} by $\rr{\hat{f}_k}{\alpha}(x)$ and sum it over all indices $k$:
\[\sum_{k}v_k(x-e_k)\rr{\hat{f}_k}{\alpha}(x) = \sum_{j}v_j(x)\left(\sum_{k}\rr{\W^x_j}{f_k}(\hat{f}(x),\alpha)\rr{\hat{f}_k}{\alpha}(x)\right) + \sum_{k}\rr{F^x}{f_k}(\hat{f}(x))\rr{\hat{f}_k}{\alpha}(x)\]
The bracket is now substituted by equation~\eqref{eq:normal:tangent}:
\[\sum_{k}v_k(x-e_k)\rr{\hat{f}_k}{\alpha}(x) = \sum_{j}v_j(x)\left(\rr{\hat{f}_j}{\alpha}(x+e_j) - \rr{\W^x_j}{\alpha}(\hat{f}(x),\alpha)\right) + \sum_{k}\rr{F^x}{f_k}(\hat{f}(x))\rr{\hat{f}_k}{\alpha}(x)\]
and both sides are summed over all lattice nodes:
\[\sum_{x,k} v_k(x-e_k)\rr{\hat{f}_k}{\alpha}(x) = \sum_{x,j} v_j(x)\rr{\hat{f}_j}{\alpha}(x+e_j) - \sum_{x,j} v_j(x)\rr{\W^x_j}{\alpha}(\hat{f}(x),\alpha) + \sum_{x,k}\rr{F^x}{f_k}(\hat{f}(x))\rr{\hat{f}_k}{\alpha}(x)\]
Finally, using Lemma~\ref{lem:tautologia}, one obtains:
\[\sum_{x,j} v_j(x)\rr{\W^x_j}{\alpha}(\hat{f}(x),\alpha) = \sum_{x,k} \rr{F^x}{f_k}(\hat{f}(x))\rr{\hat{f}_k}{\alpha}(x)=\rr{F}{\alpha}\]which concludes the proof.\end{proof}

Basing on equation~\eqref{eq:adjoint:stationary} we propose a numerical scheme in which $v(x)$ is obtained by explicit iterations:
\begin{equation}v_k^{n+1}(x-e_k) = \sum_{j}v_j^n(x)\rr{\W^x_j}{f_k}(\hat{f}(x),\alpha) + \rr{F^x}{f_k}(\hat{f}(x))\label{eq:adjoint:iteration}\end{equation}
We will call this scheme the Adjoint Lattice Boltzmann Scheme (ALBS). It has clearly the same form as the LLBS iteration formula~\eqref{eq:normal:iteration}. However the streaming of $v$ acts in the opposite direction, while the corresponding adjoint collision operator is defined as:
\begin{equation}\widehat{\W}^x_k(v,f,\alpha) = \sum_{j}v_j \rr{\W^x_j}{f_k}(f,\alpha) + \rr{F^x}{f_k}(f)\label{eq:adjoint:operator}\end{equation}

\section{Implementation of the Lattice Bolztmann Method}\label{sec:lb}
In this study a set of 19 vectors $e$ is used for 3D flow simulations~\cite{dhumieres_multiplerelaxationtime_2002}:
\begin{alignat*}{6}
e_{1} &= [0,0,0] \\
 e_{2} &= [1,0,0] \quad & e_{3} &= [-1,0,0] \quad & e_{4} &= [0,1,0] \quad & e_{5} &= [0,-1,0] \quad & e_{6} &= [0,0,1] \quad & e_{7} &= [0,0,-1] \\
 e_{8} &= [1,1,0] \quad & e_{9} &= [-1,1,0] \quad & e_{10} &= [1,-1,0] \quad & e_{11} &= [-1,-1,0] \quad & e_{12} &= [1,0,1] \quad & e_{13} &= [-1,0,1] \\
 e_{14} &= [1,0,-1] \quad & e_{15} &= [-1,0,-1] \quad & e_{16} &= [0,1,1] \quad & e_{17} &= [0,-1,1] \quad & e_{18} &= [0,1,-1] \quad & e_{19} &= [0,-1,-1]
\end{alignat*}
For the heat transfer equation we use the first $7$ of the above. For 2D cases the number of vectors is reduced to 9 for both the fluid flow and the heat transfer.

In LLMS the local collision operator plays different role at different nodes of the lattice. It is both responsible for fluid dynamics at the interior nodes, as well as for enforcing the boundary conditions. Therefore three types of nodes can be distinguished: (i)~interior nodes, (ii)~inlet/outlet nodes, (iii)~wall nodes. For inlet/outlet nodes we use simple boundary conditions by Zou and He~\cite{zou_pressure_1997}, while for the wall nodes we use bounce back condition~\cite{succi_lattice_2001}. For the interior nodes we use a Forced Multiple Relaxation Time (FMRT) collision.

\subsection{Forced Multiple Relaxation Time collision}
For a classic LB scheme, a simple (BGK) collision operator is used:
\begin{equation}\W^\text{\sc bgk}(f) = f + \omega\left(f_\eq - f\right)\label{eq:SRT}\end{equation}
where $f_\eq$ are the equilibrium densities depending on the macroscopic quantities, while $\frac{1}{\omega}$ is called the over-relaxation time. This over-relaxation time is directly related to the modelled viscosity coefficient $\frac{1}{\omega} = 0.5 + 3\nu$~($1 \leq \omega < 2$). For stabilization of the scheme an additional transformation was introduced~\cite{dhumieres_multiplerelaxationtime_2002}. This approach applies a linear transformation of the density set $f$, projecting it to a moment space (corresponding to geometric moments of the velocity set). Different relaxation factors are used in each direction of the moment space.
Let $U$ denote the matrix of this transformation. For 3D flow simulation we use the matrix proposed in~\cite{dhumieres_multiplerelaxationtime_2002}. We define the moment projection as $m = Uf$. The analogue for the BGK operator~\eqref{eq:SRT} in the moment space can be written as:
\begin{equation}\W^\text{m}(m) = m + T\left(m_\eq - m\right)\label{eq:MRT:moment}\end{equation}
where $T$ is a diagonal matrix of relaxation factors. Multiplying both sides of (\ref{eq:MRT:moment}) with $U^{-1}$ we get MRT collision
\[\W^\text{\sc mrt}(f) = f + U^{-1}TU\left(f_\eq - f\right)\]

In most cases it is less computationally expensive to calculate $m_\eq$ than $f_\eq$. This is why a intermediate form is further used:
\[\W^\text{\sc mrt}(f) = U^{-1}\Big(m_\eq + (I-T)\left(Uf - m_\eq\right)\Big)\]

For our applications it is important to change the global quantities during the iteration process in order to:
\begin{itemize}
	\item enforce certain values like velocity, temperature, etc. at specific points,
	\item optionally add a forcing term like mass force or heat source.
\end{itemize}
We introduce these changes by differentiating between pre-collision equilibrium ($f^\eq_\pre$) and post-collision ($f_\post^\eq$). This defines the Forced Multi Relaxation Time (FMRT) collision:
\begin{equation}\W^\text{\sc fmrt}(f) = U^{-1}\Big(m^\post_\eq + (I-T)\left(Uf - m^\pre_\eq\right)\Big)\end{equation}

The pre-collision equilibrium $f_\eq^\pre$ is based on global variables (like $u$ and $\rho$) of the density set $f$ prior to collision while $f_\eq^\post$ is based on desired post-collision values. The FMRT can be seen as a natural extension of the Exact Difference Method~\cite{kupershtokh_equations_2009} of discretization of body forces in LBM. Let us assume that $f^\eq$ depends on the velocity $u$ and the density $\rho$. We can apply a body force $R$ and integrate the dynamics of $u$ with a simple Euler step: $u_\post = u_\pre + R\text{dt}$. Applying this to the FMRT scheme we get:
\[\W^\text{\sc fmrt}(f) = f_\eq(u_\pre+ R\text{dt},\rho) + U^{-1}TU\left(f - f_\eq(u_\pre,\rho)\right)\]
Let us now discard the suffix of $u$ to obtain:
\[\W^\text{\sc fmrt}(f) = \Big[f_\eq(u+ R\text{dt},\rho) - f_\eq(u,\rho)\Big] + \W^\text{\sc mrt}(f)\]
where the first bracket denotes the Exact Difference discretization of the force $R$. This approach can be used for any forcing terms, including: heat flux, solvent source as well as these which enforce prescribed velocity or temperature at selected points.

\subsection{Topology parametrization}
In order to implement topology optimization with classic optimization algorithms, one has to parametrize the design space. For this purpose at each node of the lattice $L$ we define a parameter $w$, which refers to interpolation between a fluid node ($w=1$) and a solid node ($w=0$). For material properties like heat diffusivity, we use linear interpolation $\beta = w\beta_\text{fluid} + (1-w)\beta_\text{solid}$.

In order to represent different dynamics in the solid and in the fluid regions, the most common method is to add a fictitious Darcy body force $R = -K(w)u$~\cite{othmer_continuous_2008,andreasen_topology_2009}. In practical implementation we use: \[u_\post = u_\pre - \text{dt}K(w)u_\pre = G(w)u_\pre\]
where switching function $G$ assumes value $G(1)=1$ (no additional Darcy force exists in the fluid part) and $G(0)=0$ (corresponds to $u^\post = 0$ in the solid part). The particular selection of function $G$ can significantly affect the convergence of the optimization process, nevertheless this issue is not discussed here being a subject of a separate study. The common choice is a power function: $K(w)=\hat{K}w^\theta$ with $\theta=3$~\cite{andreasen_topology_2009,pingen_topology_2007}.

The final, optimized geometry should consist of either purely solid ($w=0$) or purely fluid ($w=1$) nodes. In this way a particular form of $G$ becomes unimportant for the evaluation of final result. To reduce the number of nodes with intermediate values of $w$ (between $0$ and $1$) during the optimization, we introduce a quadratic penalty function $P$:
\begin{equation}P = \sum_{nodes}w(1-w)\label{eq:P}\end{equation}
This penalty is gradually added to the main objective function to force the optimizer to avoid non-physical values of $w$.

To ensure that the final geometry has a physical meaning at a final stage of optimization, every $w$ that is below a threshold $\eta$ is set to $0$, while everything above $\eta$ is set to $1$. Finally the threshold level $\eta$ giving the best objective function value is selected from the interval $[0,1]$. Example of the dependence of the objective function on the threshold level can be observed in Fig.~\ref{pic:mixer:thres}.

\subsection{Optimization}\label{subsec:opt}
For optimization two methods are used in the present paper. The first is the Method of Moving Asymptotes (MMA) proposed by Svanberg~\cite{svanberg_method_1987}. In this approach, at each iteration of the optimization algorithm:
\begin{itemize}
\item the primal solution is iterated until convergence
\item the adjoint scheme is iterated by the same number of iterations
\item the objective function and the gradient are evaluated
\item the optimization parameters are updated by MMA
\end{itemize}

The second method is a simple descent algorithm. The optimization problem is solved simultaneously with the primal and adjoint problem using a one-shot framework~\cite{jaworski_toward_2008,kunisch_single-step_2009}. A single iteration consists of:
\begin{itemize}\item one iteration of the primal problem\item one iteration of the adjoint problem\item one update of the parameters:
\begin{equation}\label{eq:gradient:update}\alpha^{n+1}_i=\alpha^n_i + \zeta\overline{\pr{\alpha_i}F}\end{equation}\end{itemize}where $\overline{\pr{\alpha_i}F}$ denotes the present estimate of the gradient from the actual adjoint solution. The sufficiently small parameter $\zeta$ is chosen to suppress possible oscillations (caused by the delayed response).

\subsection{Programming and Parallel Implementation}
Both the primal and the adjoint Lattice Boltzmann Schemes are solved by the in-house solver CLB. This solver is highly parallel and tuned for nVidia CUDA architecture. It is also capable of running on multiple GPUs by using MPI directives.

To achieve high performance and to allow for automatic adjoint formulation, we have used special code~generation techniques. The code is generated for every model separately. The model is described by a set of tables defining:
\begin{itemize}
\item {\tt Densities} --- Names of the densities and corresponding streaming vectors (e.g. $f_i$ and $e_i$ in~\eqref{eq:normal:iteration})
\item {\tt Settings} --- Names of global settings that influence the model (e.g., $\nu$)
\item {\tt Globals} --- Names and properties of global integrals (e.g., Volume Flux, Heat Flux)
\item {\tt Quantities} --- Names of macroscopic quantities that are exported by the model (e.g., $u$, $\rho$, $T$)
\end{itemize}
All the source code is generated basing on this description and on the generic text templates. For this purpose an in-house R-Template (RT) tool is used, based on the R programing language~\cite{r_core_team_r:_2013}.

\begin{figure*}
\begin{center}
\includegraphics[width=4in]{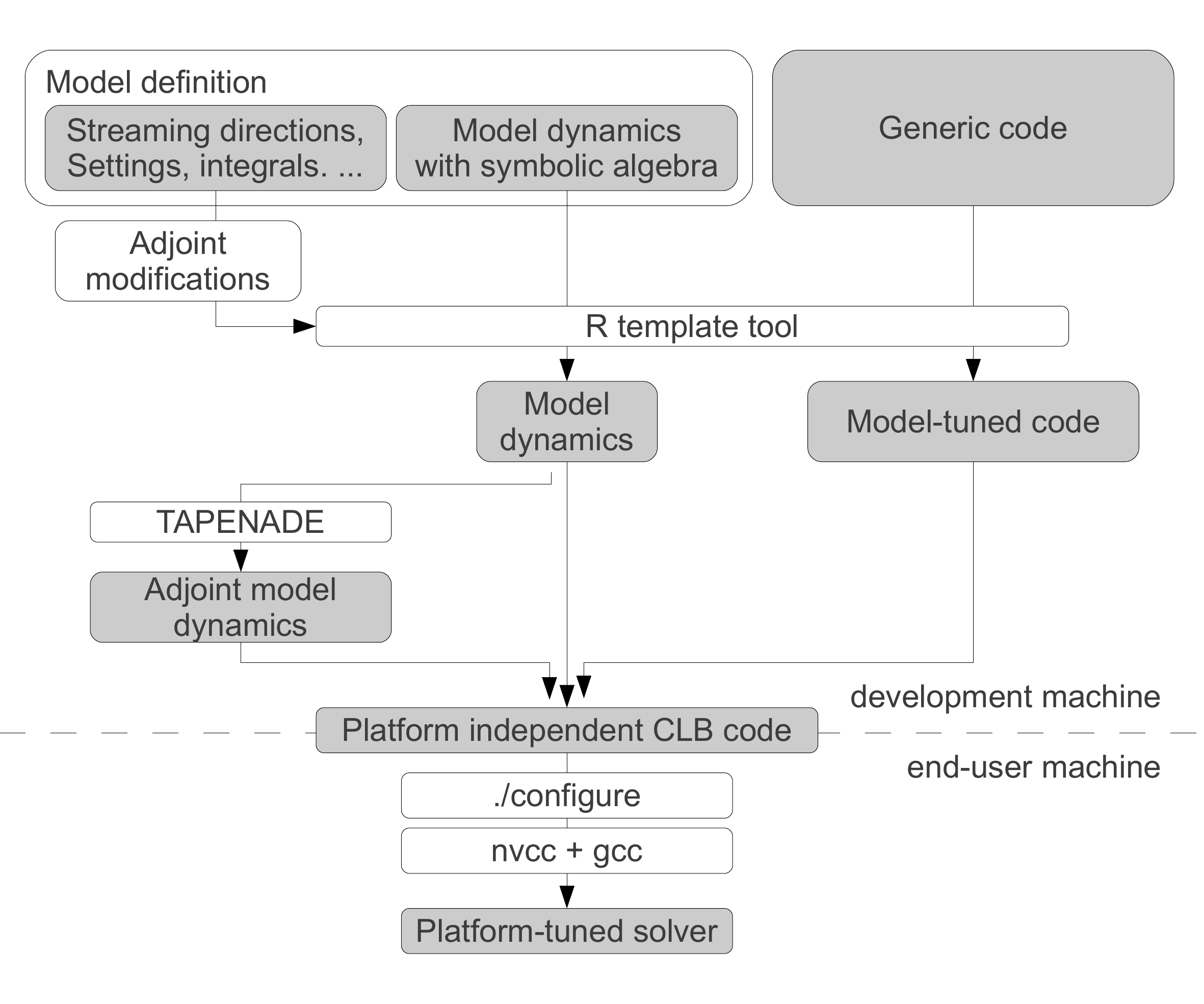}
\end{center}
\caption{Scheme of code generation and compilation in the CLB code}\label{pic:CLB:scheme}
\end{figure*}

The model collision operator $\W$ is defined as a function in a C source file, operating on the defined densities. As the code includes matrix multiplications and other repeatable operations, it is written partly using the symbolic algebra. These symbolic formulas are subsequently expanded by the R-Template tool.

For adjoint formulation, the model information is augmented with adjoint densities. Then the model's dynamics is differentiated using Automatic Differentiation (AD) tool TAPENADE~\cite{hascoet_tapenade_2013} (for reverse differentiation of operator $\W$). The produced code is subsequently used in implementation of equation~(\ref{eq:adjoint:operator}).

The complete code generation and compilation procedure is presented in Fig.~\ref{pic:CLB:scheme}.

\section{Results}\label{sec:results}
To verify the performance of the proposed optimization approach, two test cases were considered. The first test case deals with maximizing of the low Reynolds number mixing. The second test case concerns coupled heat and mass transfer, in which the convected heat is maximized. In both cases the new shapes are created in the channel with square cross-section. The topology optimization algorithm generates these shapes in a selected volume of this channel (called a design region).

The fluid flow is simulated using a standard 19-density 3D MRT~\cite{dhumieres_multiplerelaxationtime_2002}, while for the heat transfer a 7-density single relaxation time BGK is used. The temperature field is always a passive scalar and does not influence the fluid flow.

The walls of the channel are adiabatic with a no-slip boundary condition for velocity. At the inlet and at the outlet the standard Pressure Inlet/Outlet boundary conditions by Zou and He~\cite{zou_pressure_1997} are imposed. The pressure at the outlet is set to the reference level of $0$ (corresponding to the LB density $1$), while the pressure at inlet is set equal to the desired pressure drop. The additional constraint was imposed on the inlet, limiting (out of stability reasons) the velocity to $0.05$. During the optimization process in both cases the Reynolds number remains in the range of $100$--$250$, being strongly dependent on the shape of the generated geometry. Summary of settings for both cases can be found in Table~\ref{tab:cases}.

\begin{table*}[!h]
\begin{center}\footnotesize
\begin{tabular}{ccccccccc}
\toprule
Case & Lattice size & No. of nodes & DoF & \parbox{2cm}{\centering{Span of the design region}} & $\Delta p$ & $\nu$ & $\alpha_\text{fluid}$ & $\alpha_\text{solid}$\\
\midrule
Mixing & $450\times 50\times 50$ & \sci{1.12}{6} & \sci{29.2}{6} & 61--390 & 0.016666 & 0.02 & 0.003 & 0.003\\
Heat exchanger & $352\times 50\times 50$ & \sci{0.88}{6} & \sci{22.8}{6} & 101--252 & 0.03 & 0.01 & 0.003 & 1\\
\bottomrule
\end{tabular}
\end{center}
\caption{Summary of the setting in both test cases}\label{tab:cases}
\end{table*}

\subsection{Low Reynolds number mixing}
The mixing test case consists of a square channel of height $D$ and length $9D$. The design region of length $\sim 6.5D$ is located in the middle section of the channel. At the inlet discontinuous temperature distribution was prescribed ($T=-1$ for $y<25$ and $T=1$ for $y\geq 25$ --- see Fig.~\ref{pic:mixer:slice1}). As the heat conduction mimics the diffusion, the evolution of temperature can be used to represent mass mixing process. To evaluate the quality of mixing we use a flux of an award function:
\begin{equation}\text{Mixing} = \int_\text{outlet} (u\cdot n)(1-T)(1+T)dS = \int_\text{outlet} (u\cdot n)(1-T^2)dS\label{eq:mixing:obj}\end{equation}
This simple award function brings a balance between low variance of temperature and high volume of the flow. The pressure drop of $\Delta p = 0.016666$ was set between the inlet and the outlet. The viscosity was taken as $\nu=0.02$, while both the solid and the fluid heat diffusivity coefficients were set to $\alpha_\text{solid} = \alpha_\text{fluid} = 0.003$ (see Table~\ref{tab:cases}).

If there is no obstacle in the channel, the temperature profile is evolving by simple heat conduction. The resulting distribution of temperature at the outlet is illustrated in Fig.~\ref{pic:mixer:slice2}. This particular distribution has a standard deviation of $\sigma=0.761$ (see Table~\ref{tab:results}) while the value of the objective function~\eqref{eq:mixing:obj} is around $50$.

\begin{figure}[]
\begin{center}
\includegraphics[scale=1]{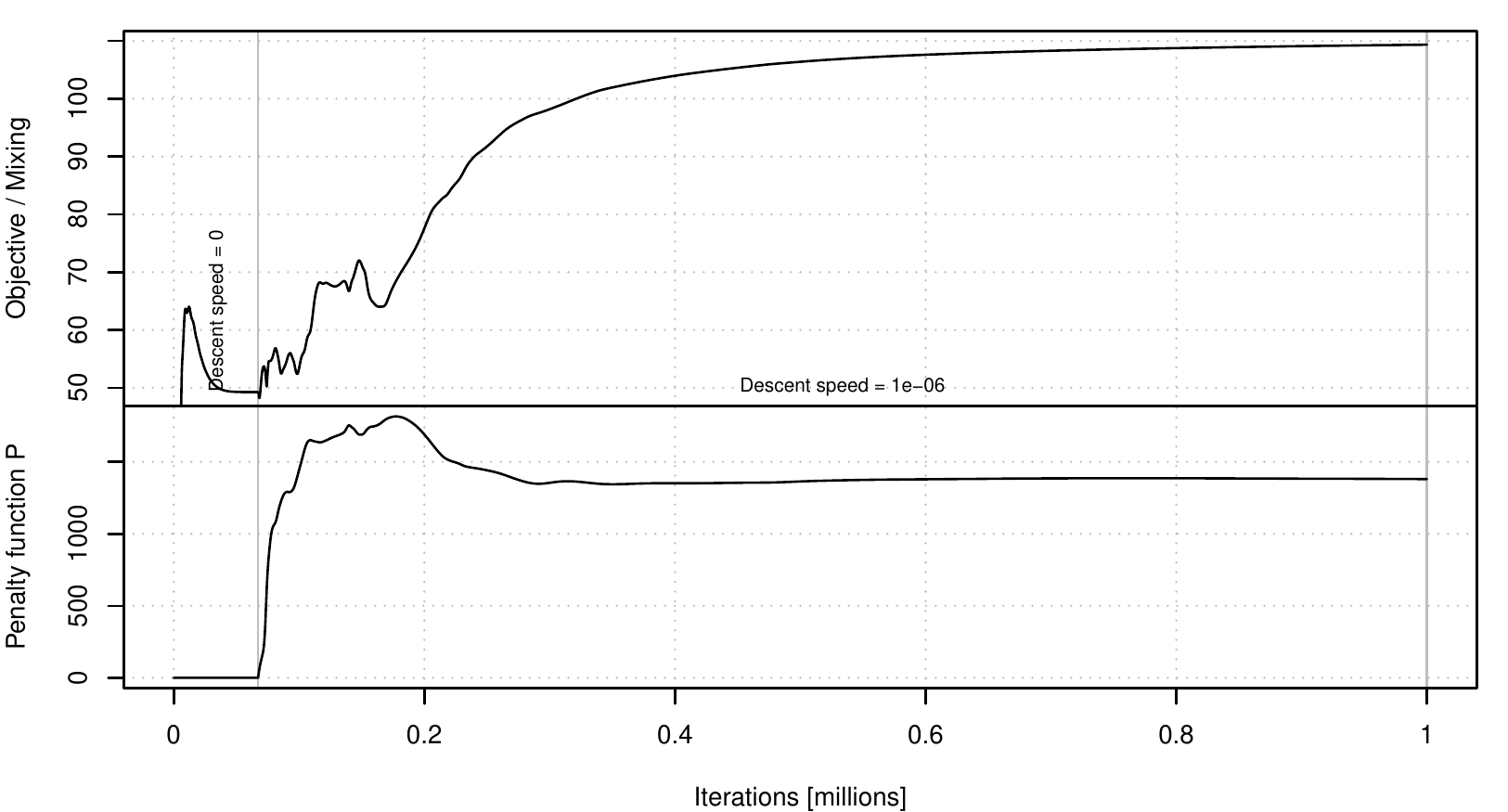}
\end{center}
\caption{Convergence of optimization of {\it free-topology} mixer. {\it Descent speed} denotes the speed $\zeta$ in the steepest descent algorithm (see~\ref{subsec:opt}).}\label{pic:mixer:conv}
\end{figure}

The convergence of the one-shot optimization procedure can be observed in Fig.~\ref{pic:mixer:conv}. The final value of the objective is around $110$ (the standard deviation of the temperature distribution is $\sigma=0.162$). The resulting geometry consists mostly of nodes with values of $w>0.9$. As the diffusivity coefficients of the solid and the fluid are the same and value of $w=0.9$ seems sufficient to completely stop the flow, the sensitivity of the objective to parameter $w$ becomes negligible.

\begin{figure}[]
\begin{center}
\includegraphics[scale=1]{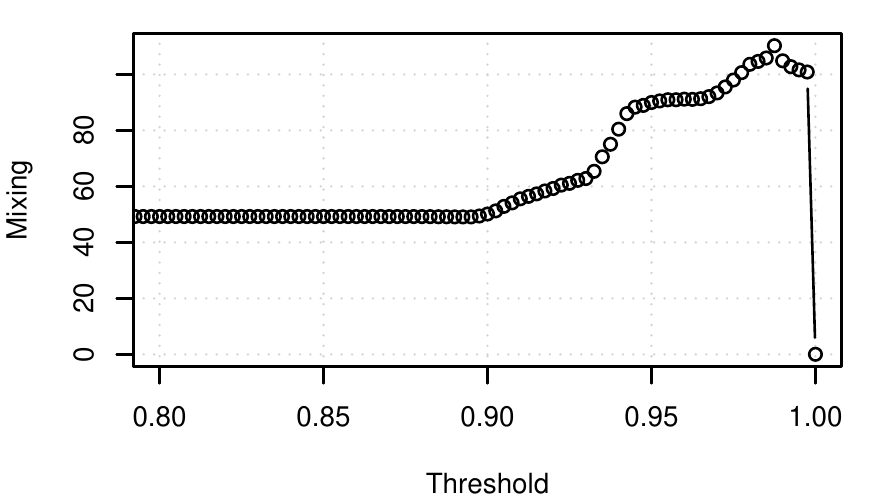}
\end{center}
\caption{Influence on the cut-off level on objective in the mixing optimization test case.}\label{pic:mixer:thres}
\end{figure}

Application of the threshold, marginally reduces the objective. Dependence of the objective on the threshold level is shown in Fig.~\ref{pic:mixer:thres}. In fact the standard deviation of the temperature is reduced now to $\sigma=0.151$, but at the same time the flow rate is reduced (see Table~\ref{tab:results}). The distribution of temperature at the outlet for the optimized geometry is illustrated in Fig.~\ref{pic:mixer:slice3}.

The MMA optimization gives just slightly better results (see Table~\ref{tab:results}). On the other hand the MMA approach needs around 6 million of iterations, whereas one-shot method needs around 1 million. This difference occurs, because in MMA optimization we iterate the solution and the adjoint until converged in each optimization iteration. In the one-shot approach we use the the imprecise adjoint solution for the update, saving a lot of iterations at the early stage of the optimization.

\begin{figure*}
\begin{center}
\includegraphics[width=7in]{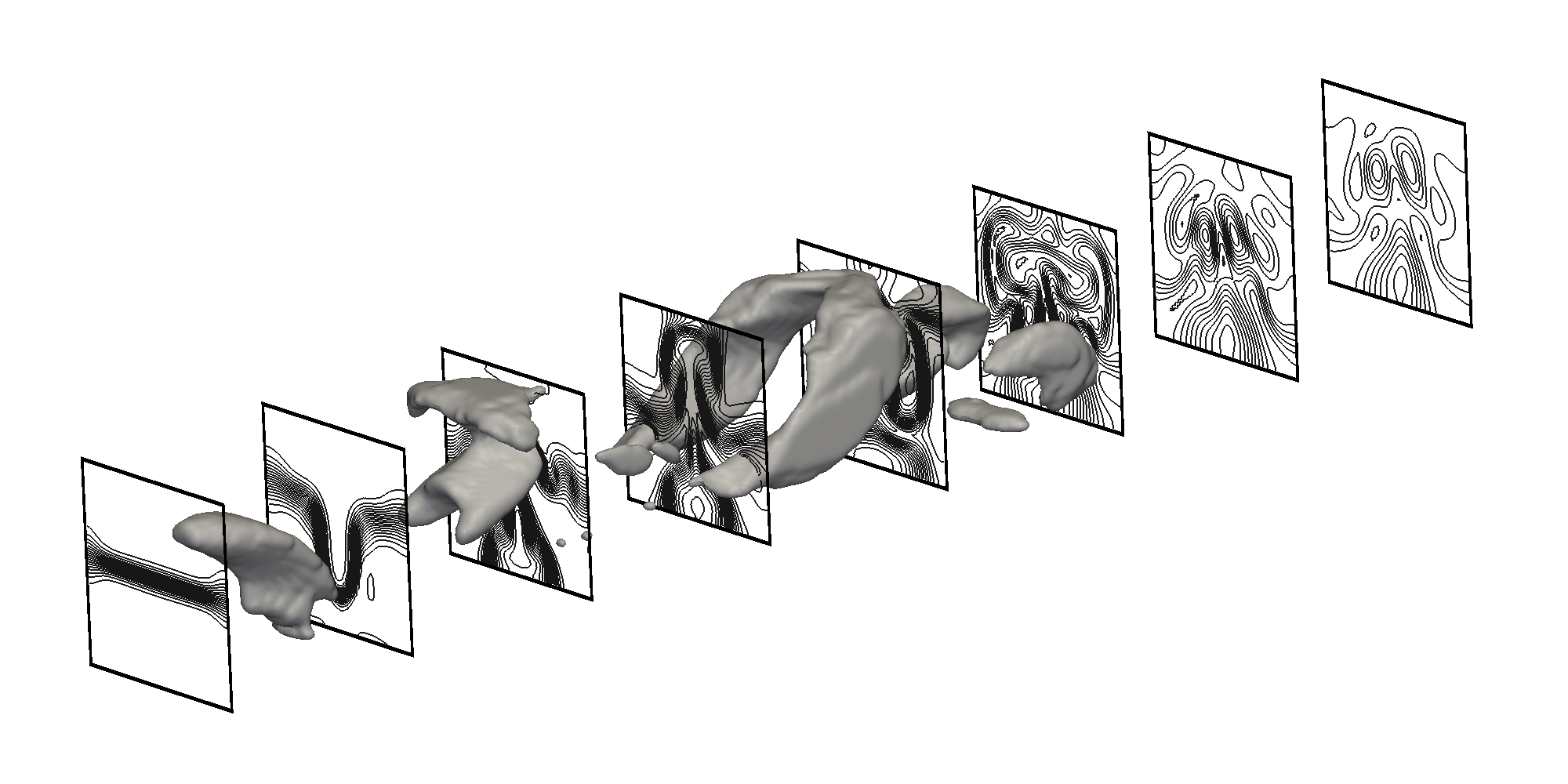}
\end{center}
\caption{Optimal free-topology mixer, generated by the one-shot algorithm. Geometry of the mixer and contour plots of temperature in the square channel.}\label{pic:mixer:geom}
\end{figure*}

The final geometry and the temperature contour plots in the channel are presented in Fig.~\ref{pic:mixer:geom}.
\newcommand{\slicesize}{2in}
\begin{figure}[]
\begin{center}
\begin{subfigure}{\slicesize}
\includegraphics[width=\slicesize]{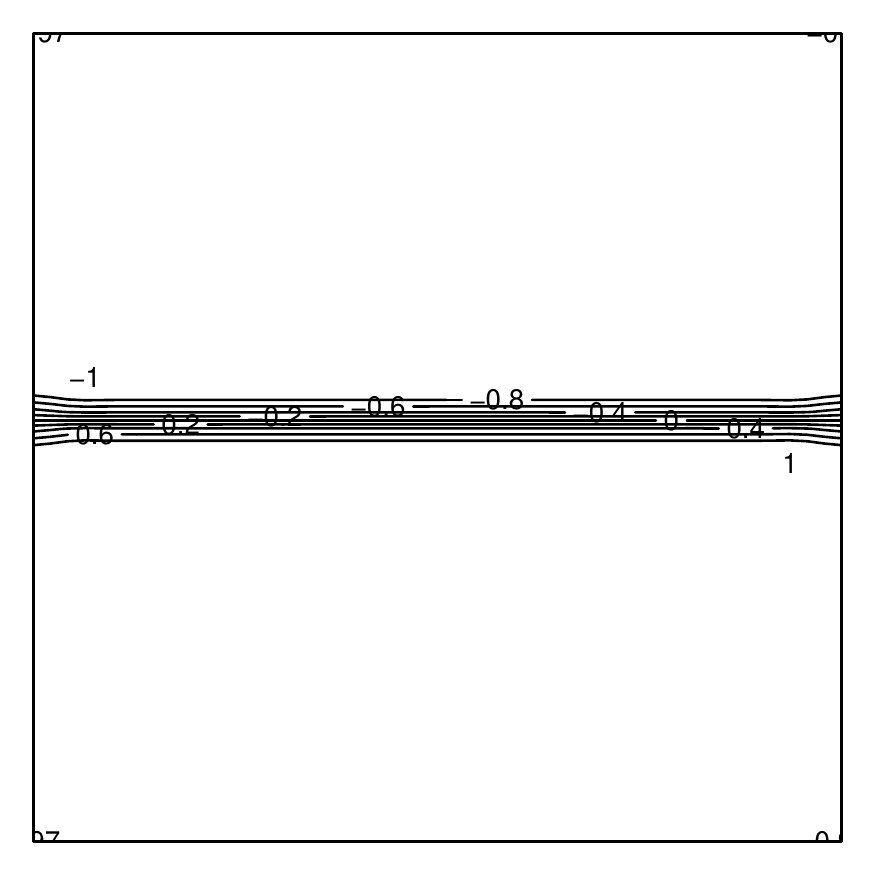}\caption{}\label{pic:mixer:slice1}
\end{subfigure}
\begin{subfigure}{\slicesize}
\includegraphics[width=\slicesize]{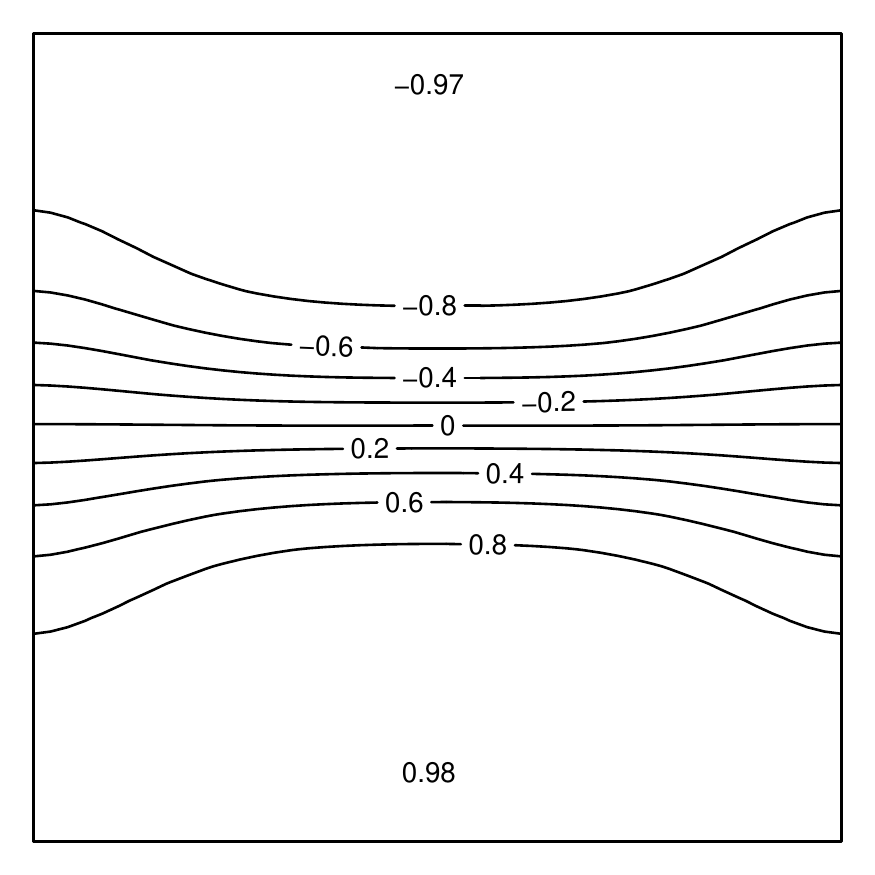}\caption{}\label{pic:mixer:slice2}
\end{subfigure}
\begin{subfigure}{\slicesize}
\includegraphics[width=\slicesize]{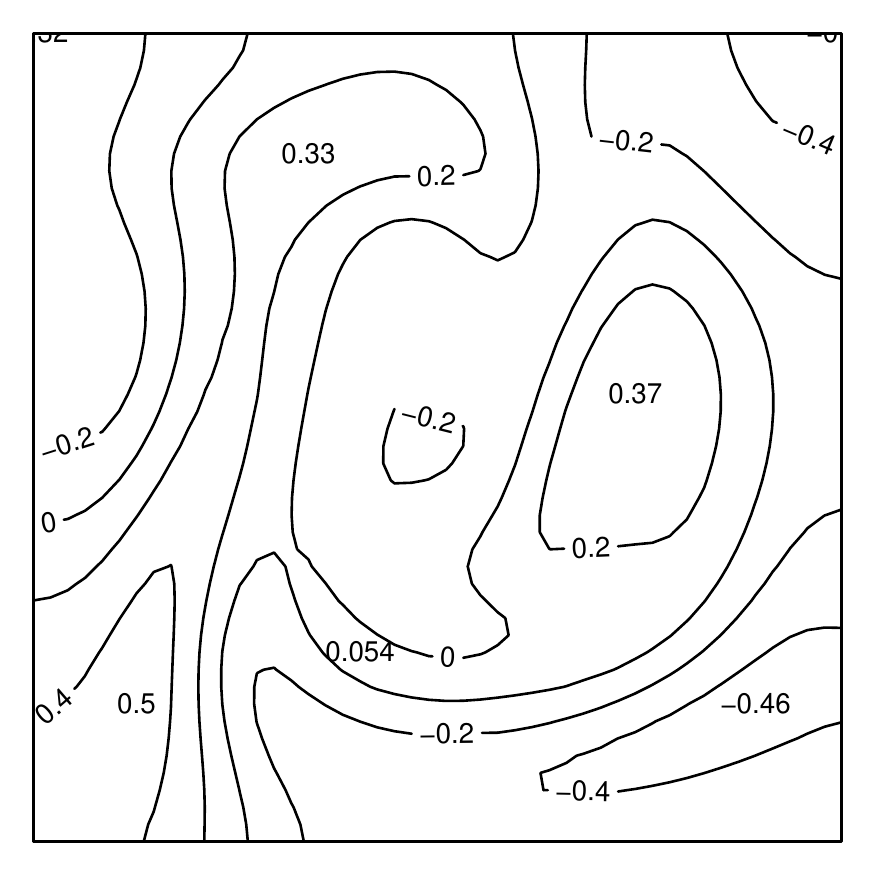}\caption{}\label{pic:mixer:slice3}
\end{subfigure}
\end{center}
\caption{Temperature on (\subref{pic:mixer:slice1}) inlet of the channel (\subref{pic:mixer:slice2}) outlet of the channel without any obstacles and (\subref{pic:mixer:slice3}) outlet of the channel with the optimal geometry.}\label{pic:mixer:slices}
\end{figure}

\subsection{Heat exchanger}
The goal of this test case is to intensify the heat exchange between a heating plate and the cooling fluid in a low Reynolds number channel flow.

\begin{figure}[h]
\begin{center}
\includegraphics[scale=1]{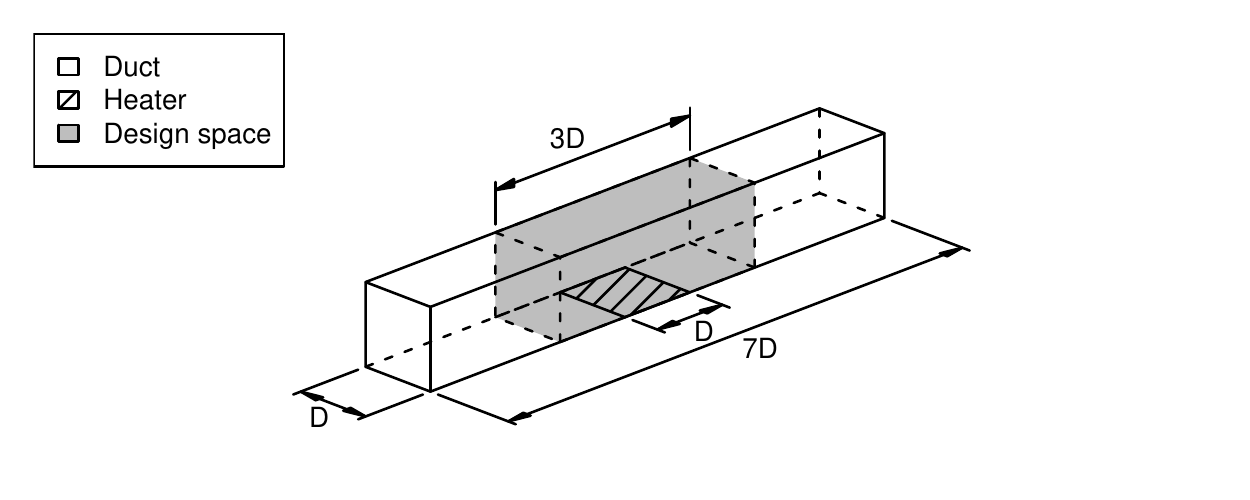}
\end{center}
\caption{Heat exchanger test case geometry}\label{pic:exchanger_geometry}
\end{figure}
The heat exchanger test case consists of a square channel of the height $D$ and the length $7D$ (See Fig.~\ref{pic:exchanger_geometry}). In the middle of the channel, a square $D\times D$ heating plate is placed at the bottom. The design region is the middle section of the pipe ($3D$). The heating plate has a fixed temperature of $T=1$, while the fluid at inlet has $T=0$. The goal is to increase the heat flux convected at the outlet.
\begin{equation}\text{Heat flux} \sim \int_\text{outlet} (u\cdot n)TdS\label{eq:exchanger:obj}\end{equation}
The pressure drop of $\Delta p = 0.03$ was set between the inlet and the outlet. The viscosity coefficient was assumed as $\nu=0.01$, while the solid and the fluid heat diffusivity coefficients were taken as $\alpha_\text{solid} = 1$ and $\alpha_\text{fluid} = 0.003$ respectively (Prandtl number $= 3.33$).

\begin{figure}[]
\begin{center}
\includegraphics[scale=1]{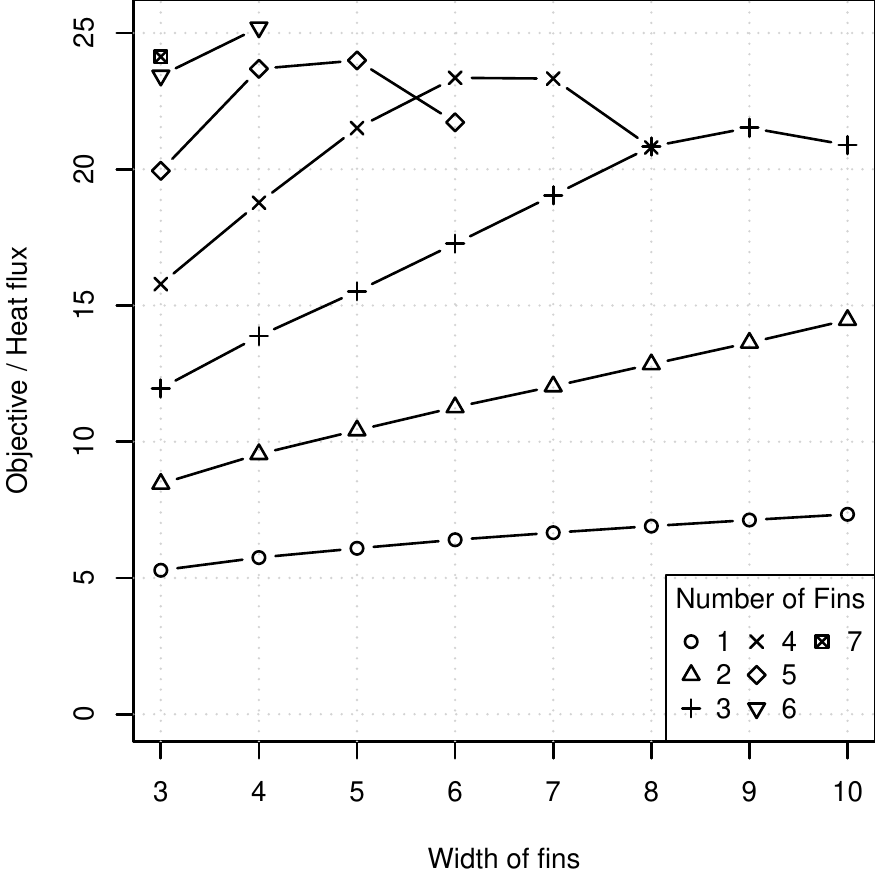}
\end{center}
\caption{Objective function for a heat exchanger consisting of an set of parallel fins.}\label{pic:exchanger_fins}
\end{figure}

If there is no obstacle in the channel, the heat exchange is marginal giving value of the objective function~\eqref{eq:exchanger:obj} of around $2.4$ and the mean temperature of $0.02$.

To make a fair comparison of the performance of the optimized geometry, a series of simple geometries were first investigated. Geometries consisting of one to seven evenly spaced parallel vertical fins of width of 3--10 lattice nodes were tested with the same flow settings. An additional constraint was adopted to preserve a gap of at least 3 lattice nodes between the fins. The results for all fin geometries can be seen in Fig.~\ref{pic:exchanger_fins}. The best heat transfer was achieved for 6 fins of width 4 (see Table~\ref{tab:results}). For such configuration the value of the objective is $25.2$.

In contrast the one-shot topology optimization algorithm gives a result of around $31$. However after application of the threshold this value reduces to $28$ (see Table~\ref{tab:results}). This behaviour is caused by a region of $w > 0.9$ acting as a porous obstacle on top of the resulting geometry. To further improve the quality of the result penalization of the parameter $w$ is used.

\begin{figure}[]
\begin{center}
\includegraphics[scale=1]{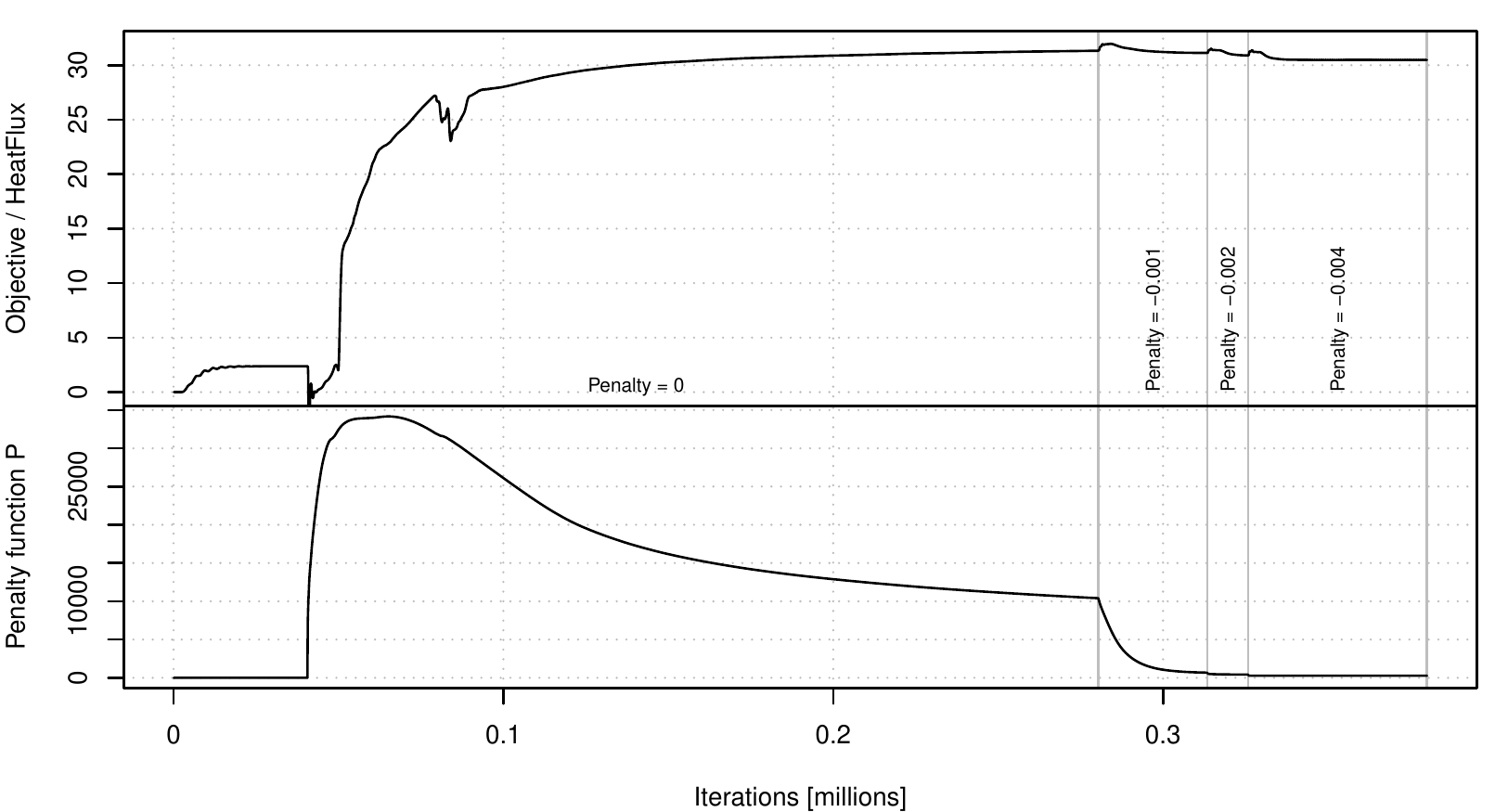}
\end{center}
\caption{Convergence of optimization of {\it free-topology} heat exchanger with the one-shot algorithm. {\it Penalty} denotes the weight of function $P$ (see~\eqref{eq:P}) in the objective function.}\label{pic:exchanger:conv}
\end{figure}

We add the function $P$~(\ref{eq:P}) to the objective, and exponentially increase the weight of this function in the overall objective. The convergence of this process can be observed in Fig.~\ref{pic:exchanger:conv}. As can be clearly seen the penalization does not reduce the true objective function significantly. On the other hand function $P$ is reduced efficiently, and the resulting geometry is more binary. Finally we loose less of the performance applying the threshold then we did in the case of optimization without penalization (compare Opt and Opt+T to Opt+P and Opt+P+T in Table~\ref{tab:results}).

\begin{figure}[]
\begin{center}
\begin{subfigure}{2.5in}
\includegraphics[width=2.5in]{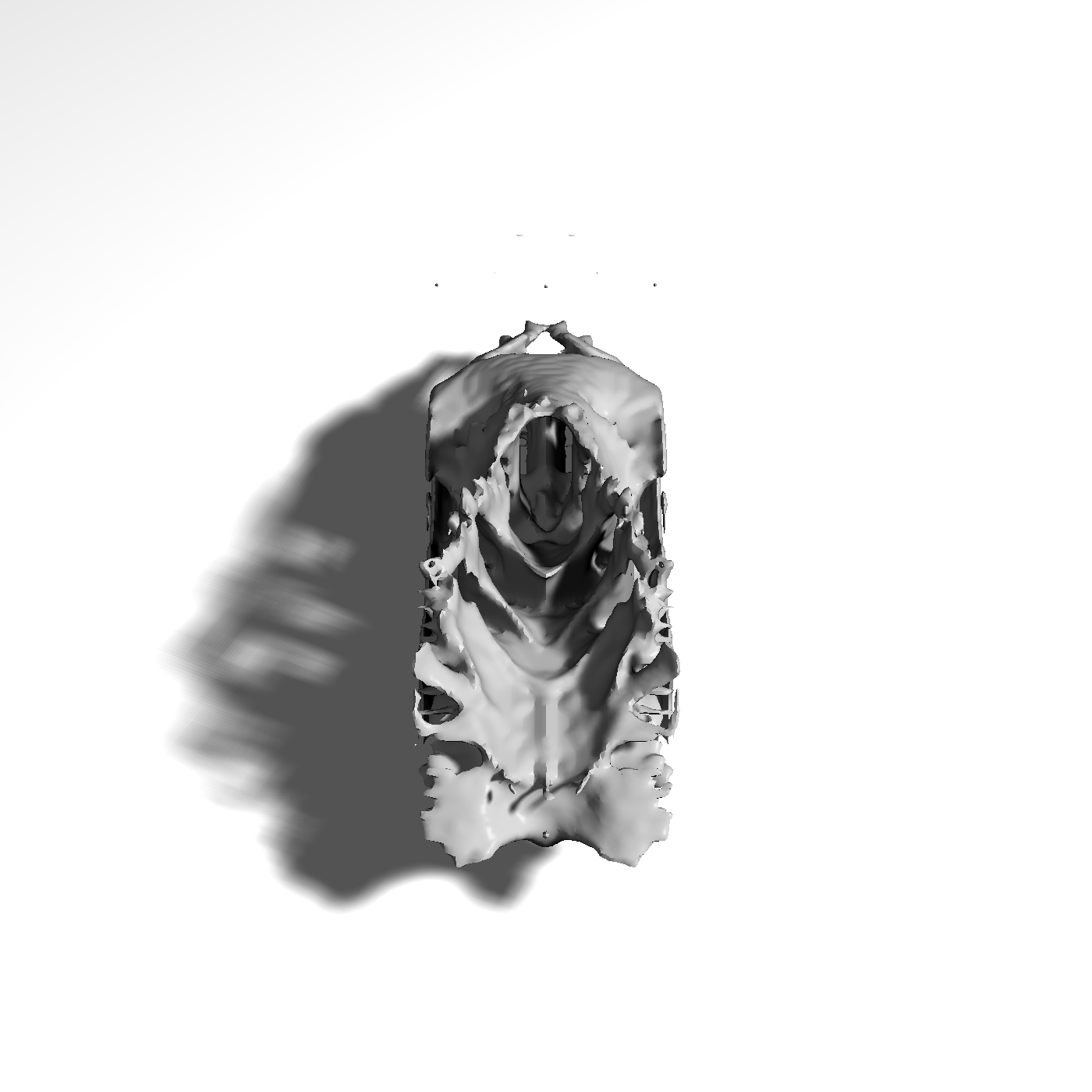}\caption{}\label{pic:exchanger:front}
\end{subfigure}
\begin{subfigure}{2.5in}
\includegraphics[width=2.5in]{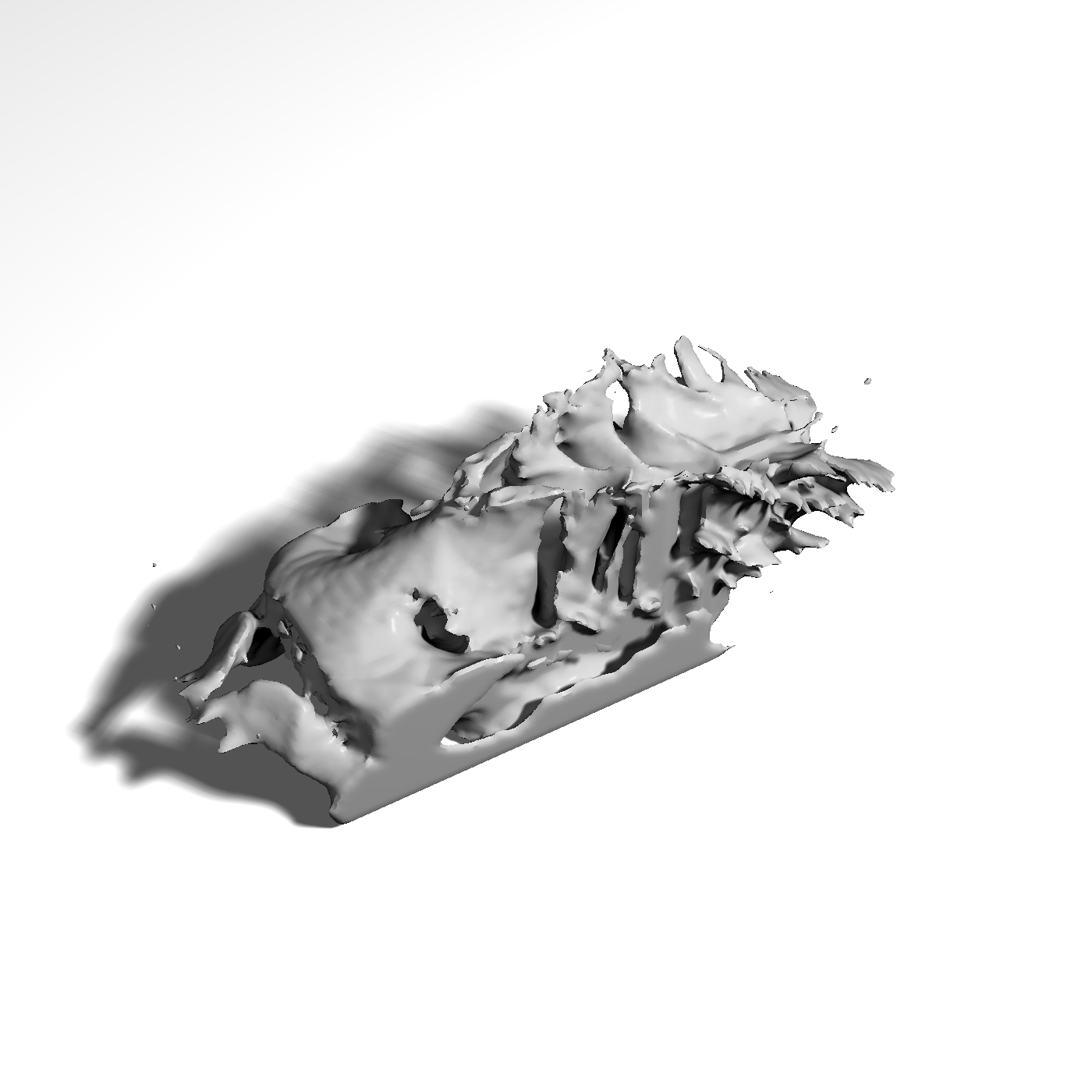}\caption{}\label{pic:exchanger:back}
\end{subfigure}
\end{center}
\caption{Geometry of {\it free-topology} heat exchanger. (\subref{pic:exchanger:front}) front and (\subref{pic:exchanger:back}) back.}\label{pic:exchanger:geometry}
\end{figure}

The optimized geometry is shown in Figure~\ref{pic:exchanger:geometry}. The increase in the objective function is $17\%$ and in the mean temperature $21\%$, compared to the best fin geometry. When compared to the empty channel, the objective is $12$ times higher, and the mean temperature $22$ times higher.

\begin{figure}[]
\begin{center}
\includegraphics[scale=1]{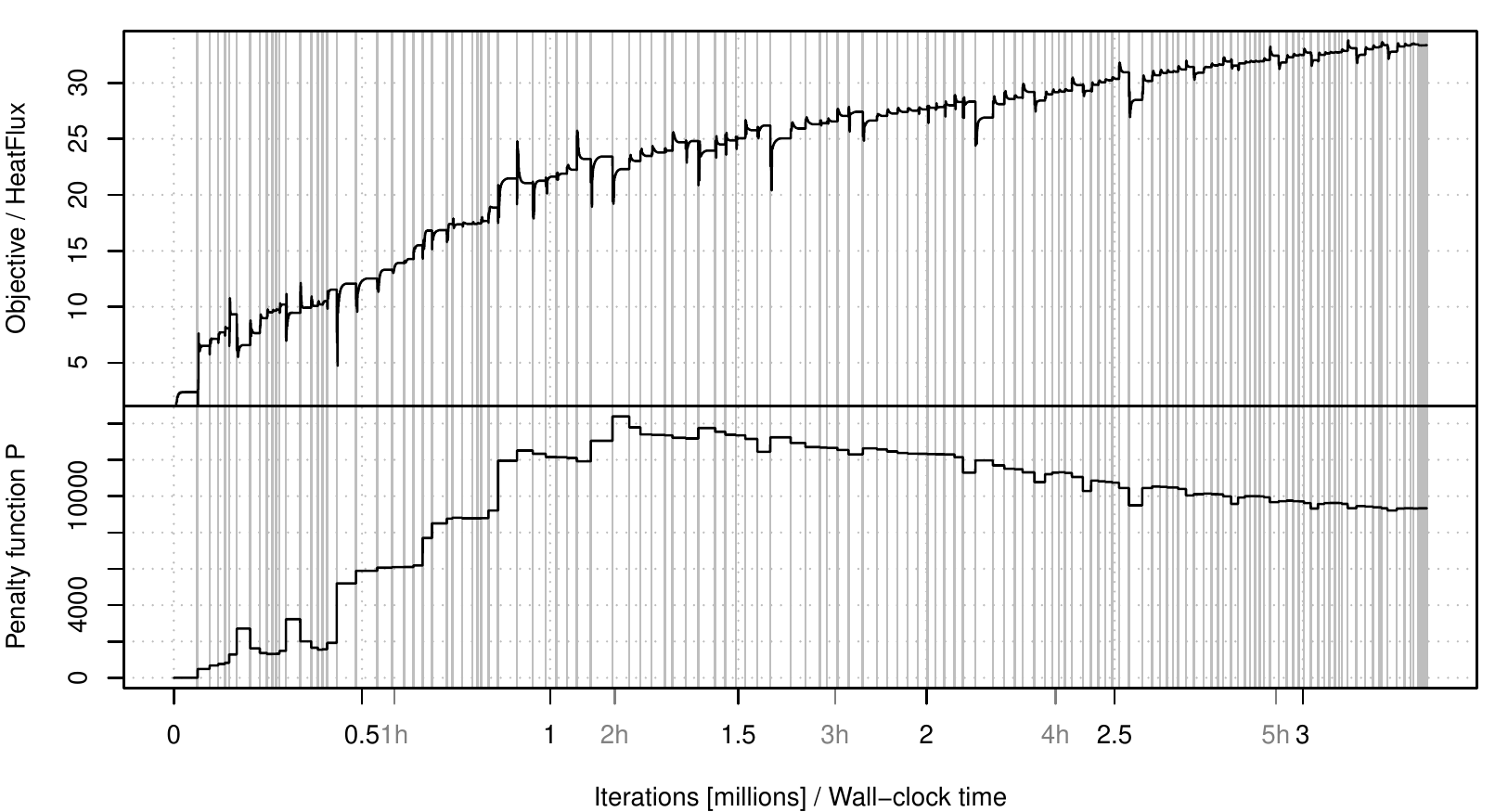}
\end{center}
\caption{Convergence of optimization of {\it free-topology} heat exchanger with MMA algorithm. In each iteration of MMA algorithm (separated by vertical lines) the solution of the primal and adjoint problem is iterated until converged (see~\ref{subsec:opt}).}\label{pic:exchanger:convMMA}
\end{figure}

Using the MMA optimization provides us with even better result of the objective function around $33$, but is reduced to $27.8$ after applying the threshold. As in the earlier case, the one-shot optimization outperforms the MMA approach. Convergence of MMA optimization is shown in Fig.~\ref{pic:exchanger:convMMA}. The number of iterations needed for a convergence of the solution (without the optimization) is around 40 thousand, for one-shot optimization (with penalization) it is 0.4 million, and for MMA optimization it is around 3.5 million. Additionally there is no simple method of adding the penalty gradually in MMA, which could prevent the drop of the objective function caused by the application of the threshold on the geometry.

\newcommand{\myinterval}[1]{\multicolumn{1}{c}{\hspace{-0.1cm}{$#1$}\hspace{-0.1cm}}}
\begin{table*}
\caption{Summary of the results. SD --- Standard deviation; Opt --- Optimized with one-shot method; MMA --- Optimized with MMA algorithm; T --- threshold applied, resulting in a binary ($w=0$ and $w=1$) geometry; P --- penalty $P$ (see~\eqref{eq:P}) gradually added to the objective function; Fins --- Optimal fin heat exchanger}\label{tab:results}
\begin{center}\footnotesize
\begin{tabular}{rlrrrrrrrrrrrrr}
\toprule
& \multirow{3}{*}{Case} & \multicolumn{1}{c}{\multirow{3}{*}{Objective}} & \multicolumn{2}{c}{Velocity} & \multicolumn{2}{c}{Temperature}&\multicolumn{5}{c}{Topology parameter $w$ in design region}&\multicolumn{1}{c}{\multirow{2}{*}{Iterations}}\\
\cmidrule(r){4-5}\cmidrule(r){6-7}\cmidrule(r){8-12}
& & & \multicolumn{1}{c}{Mean} &
\multicolumn{1}{c}{SD} &
\multicolumn{1}{c}{Mean} &
\multicolumn{1}{c}{SD} &
\multicolumn{1}{c}{0} &
\myinterval{(0,0.9)} & \myinterval{[0.9,0.99)} & \myinterval{[0.99,1)} & \multicolumn{1}{c}{1} \\
\midrule\multirow{5}{*}{\rotatebox{90}{Mixer}} & Empty & 49.266 & 0.051 & 0.031 & 0.000 & 0.763 & --- & --- & --- & --- & 100\% & \sci{0.04}{6}\\
 & MMA & 114.508 & 0.050 & 0.030 & 0.041 & 0.068 & $<$1\% &  1\% &  6\% &  6\% & 87\% & \sci{6}{6} \\
 & MMA+T & 111.863 & 0.050 & 0.030 & 0.041 & 0.175 &  2\% & --- & --- & --- & 98\% & ---\\
 & Opt & 113.381 & 0.050 & 0.031 & 0.000 & 0.094 & --- & --- &  6\% &  7\% & 87\% & \sci{1}{6}\\
 & Opt+T & 110.383 & 0.050 & 0.030 & 0.000 & 0.200 &  3\% & --- & --- & --- & 97\% & ---\\
\midrule\multirow{8}{*}{\rotatebox{90}{Exchanger}} & Empty & 2.376 & 0.051 & 0.029 & 0.020 & 0.049 & --- & --- & --- & --- & 100\% & \sci{0.03}{6}\\
 & Fins & 25.202 & 0.030 & 0.017 & 0.367 & 0.092 & 16\% & --- & --- & --- & 84\% & \sci{0.03}{6}\\
 & MMA & 33.362 & 0.034 & 0.020 & 0.424 & 0.101 & 10\% & 14\% &  7\% &  4\% & 65\% & \sci{3.5}{6}\\
 & MMA+T & 27.883 & 0.032 & 0.017 & 0.378 & 0.062 & 25\% & --- & --- & --- & 75\% & ---\\
 & Opt & 31.032 & 0.025 & 0.009 & 0.533 & 0.050 & 25\% & 15\% &  3\% &  9\% & 48\% & \sci{0.4}{6}\\
 & Opt+T & 28.226 & 0.026 & 0.013 & 0.467 & 0.065 & 40\% & --- & --- & --- & 60\% & ---\\
 & Opt+P & 30.955 & 0.027 & 0.010 & 0.502 & 0.051 & 31\% & $<$1\% &  2\% &  9\% & 57\% & \sci{0.4}{6}\\
 & Opt+P+T & 29.447 & 0.029 & 0.014 & 0.446 & 0.060 & 32\% & --- & --- & --- & 68\% & ---\\
\bottomrule
\end{tabular}
\end{center}
\end{table*}

\subsection{Parallel performance}

\begin{figure}[]
\begin{center}
\includegraphics[scale=1]{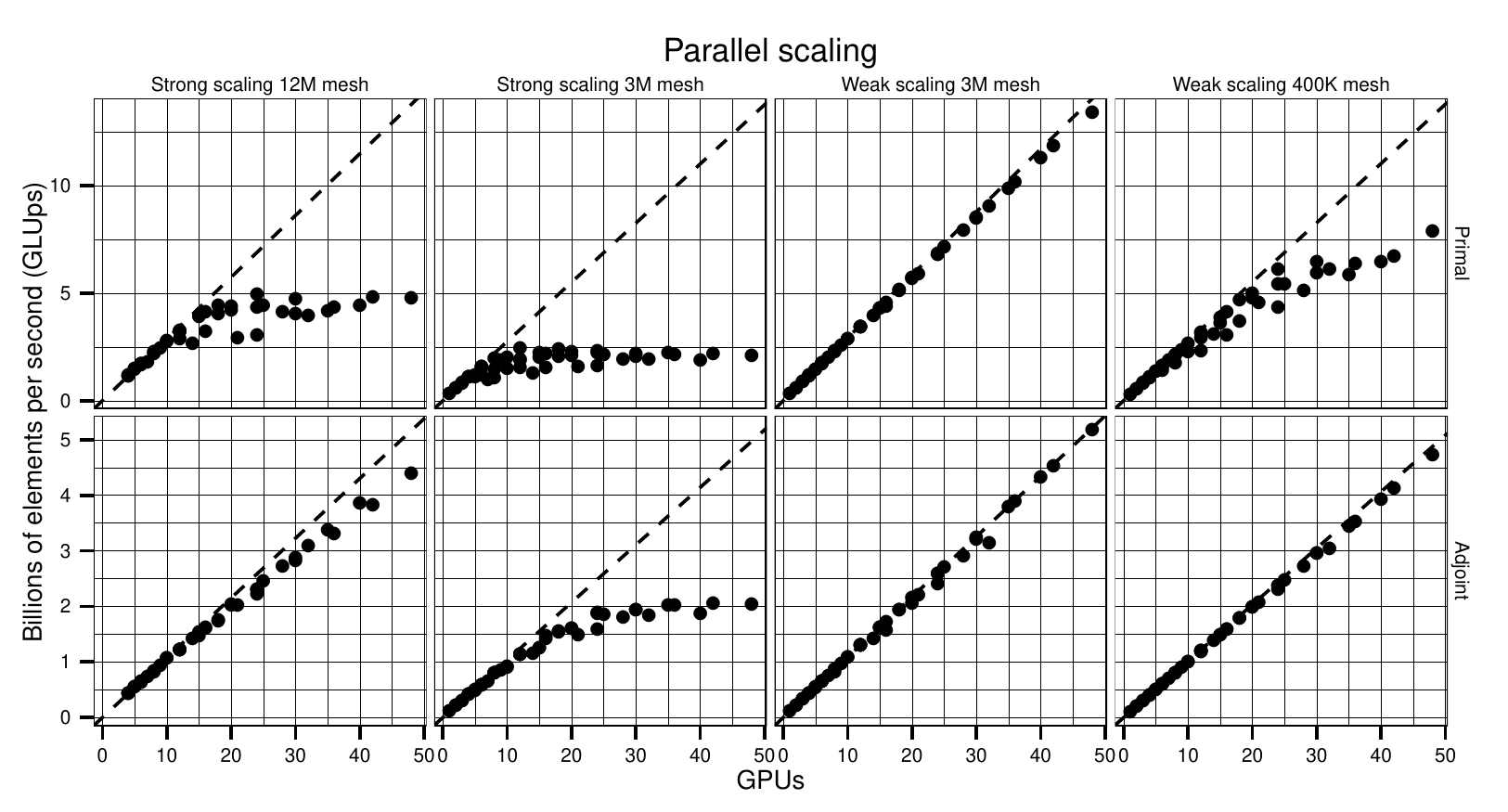}
\end{center}
\caption{Weak and strong scaling of Primal and Adjoint calculations in CLB code.}\label{pic:parallel:scaling}
\end{figure}

Parallel performance of the final code was investigated in a series of numerical experiments. The performance of any parallel code is strongly dependent on the size of the problem and the ratio between the calculations and communication. The latter however is directly related to the ratio of volume to surface of the part of the problem that each processor/GPU is calculating. We distinguish two type of scaling: weak and strong. Weak scaling measures speedup of the code, if the size of the subproblem assigned to each processor is the same and fixed. Strong scaling on the other hand measures the speedup with a fixed size of the overall problem --- causing the problem to be divided to smaller parts for higher number of processors. In most cases, as the size of the subproblems decreases, the communication starts to dominate the calculations. Such system reaches a level at which adding new processors/GPUs, doesn't improve the performance. Such level is sometimes called saturation. We selected three meshes:
\begin{itemize}
\item ''3M'' --- $320\times 100\times 100$ --- \sci{3.2}{6} elements --- \sci{83}{6} DoF
\item ''400K'' --- $160\times 50\times 50$ --- \sci{400}{3} elements --- \sci{10}{6} DoF
\item ''12M'' --- $320\times 200\times 200$ --- \sci{12.8}{6} elements --- \sci{332}{6} DoF
\end{itemize}

A series of calculations were done on 1--6 nodes with 1--8 GPUs per node (1--48 GPUs in total). Calculations were carried out on nodes of ZEUS supercomputer at the Cyfronet computing center. Each computational node of the cluster is equipped with 2 Intel Xeon E5645 processors and 8 Nvidia Tesla M2090 GPU units. Weak scaling was investigated for ''3M'' and ''400K'' meshes, by running the CLB code for a mesh size proportional to the number of GPUs (in this case \sci{3.2}{6} and \sci{400}{3} elements per GPU). Strong scaling was checked for ''3M'' and ''12M'' meshes, by running the CLB code, for a fixed size mesh. The results are in Table~\ref{tab:parallel:scaling} and in Fig.~\ref{pic:parallel:scaling}. All calculations were performed in a single precision. The size of ''12M'' mesh is slightly above the upper limit for simulation of fluid with temperature and adjoint on a single Nvidia Tesla M2090 (6 GB GDDR5).

The results clearly demonstrate that the code has nearly linear weak scaling for \sci{3.2}{6} elements per node. Strong scaling saturates at 15 GPUs for ''12M'' mesh, and at 10 GPUs for ''3M'' mesh. The interesting feature of good parallelism of the proposed adjoint formulation is that the adjoint calculations are closer to the linear scaling, than the primal calculations. This is caused by a average higher number of operations per node, while keeping the communication at exactly the same level. The speed ratio between adjoint and primal is consistently around $2.65$, which is in good agreement with theoretical estimates of 2--3 (as derivative of one multiplication is two multiplications and one addition).

\begin{table*}
\caption{Summary of performance tests.}\label{tab:parallel:scaling}
\begin{center}\footnotesize
\begin{tabular}{crrrrrrrr}
\toprule
\multirow{4}{*}{\rotatebox{90}{GPUs}} & \multicolumn{4}{c}{Strong scaling} & \multicolumn{4}{c}{Weak scaling} \\
\cmidrule(r){2-5}\cmidrule(r){6-9}
& \multicolumn{2}{c}{12M mesh} & \multicolumn{2}{c}{3M mesh} & \multicolumn{2}{c}{3M mesh} & \multicolumn{2}{c}{400K mesh} \\
\cmidrule(r){2-3}\cmidrule(r){4-5}\cmidrule(r){6-7}\cmidrule(r){8-9}
& \multicolumn{1}{c}{Primal} & \multicolumn{1}{c}{Adjoint} & \multicolumn{1}{c}{Primal} & \multicolumn{1}{c}{Adjoint} & \multicolumn{1}{c}{Primal} & \multicolumn{1}{c}{Adjoint} & \multicolumn{1}{c}{Primal} & \multicolumn{1}{c}{Adjoint} \\
\midrule
1 & --- & --- &  298 (---) &  109 (---) &  298 (---) &  109 (---) &  289 (---) &  104 (---) \\
2 &  604 (---) &  219 (---) &  582 ( 2\%) &  211 ( 4\%) &  585 ( 2\%) &  217 ($<1\%$) &  549 ( 5\%) &  203 ( 2\%) \\
3 &  871 (---) &  324 ($<$1\%) &  837 ( 6\%) &  305 ( 7\%) &  874 ( 2\%) &  325 ($<$1\%) &  819 ( 6\%) &  304 ( 2\%) \\
4 & 1162 (---) &  432 (---) & 1089 ( 9\%) &  411 ( 6\%) & 1161 ( 3\%) &  432 ( 1\%) & 1079 ( 7\%) &  403 ( 3\%) \\
5 & 1471 (---) &  546 (---) & 1184 (21\%) &  493 (10\%) & 1455 ( 2\%) &  542 ($<$1\%) & 1364 ( 6\%) &  506 ( 2\%) \\
6 & 1692 ( 3\%) &  641 ( 1\%) & 1557 (13\%) &  588 (10\%) & 1735 ( 3\%) &  649 ( 1\%) & 1601 ( 8\%) &  601 ( 3\%) \\
7 & 1787 (12\%) &  733 ( 3\%) &  944 (55\%) &  646 (16\%) & 2018 ( 3\%) &  750 ( 2\%) & 1860 ( 8\%) &  701 ( 3\%) \\
8 & 2271 ( 2\%) &  830 ( 4\%) & 1957 (18\%) &  804 ( 8\%) & 2307 ( 3\%) &  865 ( 1\%) & 2110 ( 9\%) &  797 ( 4\%) \\
9 & 2418 ( 7\%) &  926 ( 5\%) & 1880 (30\%) &  849 (14\%) & 2577 ( 4\%) &  971 ( 1\%) & 2365 ( 9\%) &  897 ( 4\%) \\
10 & 2797 ( 4\%) & 1063 ( 2\%) & 2014 (33\%) &  910 (17\%) & 2874 ( 4\%) & 1078 ( 1\%) & 2649 ( 8\%) &  997 ( 4\%) \\
12 & 3235 ( 7\%) & 1215 ( 6\%) & 2452 (32\%) & 1140 (13\%) & 3442 ( 4\%) & 1297 ( 1\%) & 3181 ( 8\%) & 1194 ( 4\%) \\
14 & 2639 (35\%) & 1415 ( 7\%) & 1256 (70\%) & 1148 (25\%) & 3952 ( 5\%) & 1420 ( 7\%) & 3081 (24\%) & 1384 ( 5\%) \\
15 & 3963 ( 9\%) & 1541 ( 5\%) & 2214 (51\%) & 1252 (24\%) & 4293 ( 4\%) & 1620 ( 1\%) & 3869 (11\%) & 1490 ( 4\%) \\
16 & 4105 (12\%) & 1614 ( 7\%) & 2186 (54\%) & 1468 (16\%) & 4561 ( 4\%) & 1724 ( 1\%) & 4123 (11\%) & 1583 ( 5\%) \\
18 & 4427 (15\%) & 1748 (10\%) & 2394 (55\%) & 1544 (22\%) & 5148 ( 4\%) & 1941 ( 1\%) & 4657 (11\%) & 1788 ( 4\%) \\
20 & 4384 (25\%) & 2034 ( 6\%) & 2258 (62\%) & 1605 (27\%) & 5716 ( 4\%) & 2157 ( 1\%) & 4976 (14\%) & 1981 ( 4\%) \\
21 & 2907 (52\%) & 2020 (11\%) & 1577 (75\%) & 1489 (35\%) & 5902 ( 6\%) & 2203 ( 4\%) & 4559 (25\%) & 2075 ( 5\%) \\
24 & 4943 (29\%) & 2309 (11\%) & 2292 (68\%) & 1887 (28\%) & 6855 ( 4\%) & 2585 ( 1\%) & 6087 (12\%) & 2369 ( 5\%) \\
25 & 4406 (39\%) & 2458 ( 9\%) & 2126 (71\%) & 1849 (32\%) & 7123 ( 4\%) & 2696 ( 1\%) & 5390 (25\%) & 2468 ( 5\%) \\
28 & 4124 (49\%) & 2718 (10\%) & 1916 (77\%) & 1800 (41\%) & 7910 ( 5\%) & 2905 ( 5\%) & 5127 (37\%) & 2720 ( 6\%) \\
30 & 4725 (46\%) & 2865 (12\%) & 2185 (76\%) & 1931 (41\%) & 8527 ( 5\%) & 3235 ( 1\%) & 6454 (26\%) & 2960 ( 5\%) \\
32 & 3952 (57\%) & 3088 (11\%) & 1933 (80\%) & 1838 (48\%) & 9051 ( 5\%) & 3136 (10\%) & 6088 (34\%) & 3040 ( 8\%) \\
35 & 4178 (59\%) & 3362 (11\%) & 2232 (79\%) & 2022 (47\%) & 9866 ( 5\%) & 3782 ( 1\%) & 5828 (42\%) & 3439 ( 5\%) \\
36 & 4346 (58\%) & 3306 (15\%) & 2115 (80\%) & 2009 (49\%) & 10169 ( 5\%) & 3889 ( 1\%) & 6380 (39\%) & 3526 ( 6\%) \\
40 & 4408 (62\%) & 3858 (11\%) & 1889 (84\%) & 1869 (57\%) & 11296 ( 5\%) & 4322 ( 1\%) & 6433 (44\%) & 3924 ( 5\%) \\
42 & 4790 (61\%) & 3820 (16\%) & 2158 (83\%) & 2048 (55\%) & 11850 ( 5\%) & 4528 ( 1\%) & 6685 (45\%) & 4124 ( 5\%) \\
48 & 4743 (66\%) & 4388 (15\%) & 2080 (85\%) & 2038 (61\%) & 13385 ( 6\%) & 5169 ( 1\%) & 7876 (43\%) & 4718 ( 5\%) \\
\bottomrule
\end{tabular}
\end{center}
\end{table*}

\section{Conclusions}\label{sec:conclusions}

In the present work we showed an efficient technique for addressing topology optimization in complex mesoscopic flow problems. This approach is based on solving the problem with Lattice Boltzmann Method, calculation of the sensitivity with Adjoint Lattice Boltzmann, and optimization with one-shot steepest descent method, or MMA.

This approach was successfully used for optimization of free-topology mixer and heat exchanger (see Table \ref{tab:results}).

In the present work we showed an efficient technique for addressing topology optimization in complex
mesoscopic flow problems. This approach is based on solving the problem using Lattice Boltzmann Method,
while the sensitivities are evaluated by the 
 Adjoint Lattice Boltzmann. The optimization process is carried out with one-shot steepest
descent method, or the Method of Moving Asymptotes.
This approach was successfully used for optimization of a free-topology mixer and a heat exchanger (see
Table~\ref{tab:results}). The resulting optimal geometry has a very complex, almost fractal shape.

It was also demonstrated that the code has nearly linear weak scaling, while the
strong scaling saturates at 10-15 GPUs. This was made possible by application of
various code generation and parallel programming techniques. Proposed adjoint problem formulation has even better parallel properties, keeping linear scalability even for smaller meshes.

Overall, we found Adjoint Lattice Bolzmann to be a powerful technique for topology optimization in mesoscopic multi-physics flow problems. Using this technique for inverse design, optimal control and time-dependent problems, needs further research.

\section{Acknowledgements}
This work was supported by MNiSW grant \emph{Properties and design of structures with open porosity
}(Nr MNiSW: N N507 273636) and also supported by PL-Grid Infrastructure. Most of the calculations were performed on ZEUS cluster at Cyfronet computing center.

\bibliographystyle{elsarticle-num}
\bibliography{../../main}{}

\end{document}